\documentclass[journal]{IEEEtran}
\usepackage{graphicx}
\usepackage{cite}
\usepackage{subfigure}
\usepackage{xspace}
\usepackage{amsmath}

\usepackage{hyperref}
\usepackage{multirow}
\usepackage[all]{hypcap} % fix links to figs/tabs

\newcommand{\matriplex}{\textsc{Matriplex}\xspace}
\newcommand{\cplusplus}{{\small\textsc{C++}}\xspace}
 
\begin{document}
\title{Kalman-Filter-Based Particle Tracking on Parallel
Architectures at Hadron Colliders}
% Author list
\author{G.~Cerati, M.~Tadel, F.~W\"urthwein, A.~Yagil, S.~Lantz,
  K.~McDermott, D.~Riley, P.~Wittich and  P.~Elmer.
\thanks{Manuscript received December 7, 2015.
This work was supported by the National Science Foundation.}% <-this % stops a space
\thanks{G.~Cerati, M.~Tadel, F.~W\"urthwein, A.~Yagil: UC San Diego.}%
\thanks{S.~Lantz, K.~McDermott, D.~Riley and P.~Wittich: Cornell University.}%
\thanks{P.~Elmer: Princeton University.}% 
}
\maketitle
% after \maketitle, add the following command to remove the header on the title page
\thispagestyle{empty}
% abstract usually in passive voice
\begin{abstract}
  Power density constraints are limiting the performance improvements
  of modern CPUs. To address this we have seen the introduction of
  lower-power, multi-core processors such as GPGPU, ARM and Intel
  MIC. To stay within the power density limits but still obtain
  Moore's Law performance/price gains, it will be necessary to
  parallelize algorithms to exploit larger numbers of lightweight
  cores and specialized functions like large vector units. Track
  finding and fitting is one of the most computationally challenging
  problems for event reconstruction in particle physics. At the
  High-Luminosity Large Hadron Collider (HL-LHC), for example, this
  will be by far the dominant problem. The need for greater
  parallelism has driven investigations of very different track
  finding techniques such as Cellular Automata or Hough
  Transforms. The most common track finding techniques in use today,
  however, are those based on the Kalman Filter. Significant
  experience has been accumulated with these techniques on real
  tracking detector systems, both in the trigger and offline. They are
  known to provide high physics performance, are robust, and are in
  use today at the LHC. We report on porting these algorithms to new
  parallel architectures.  Our previous investigations showed that,
  using optimized data structures, track fitting with Kalman Filter
  can achieve large speedups both with Intel Xeon and Xeon Phi. We
  report here our progress towards an end-to-end track reconstruction
  algorithm fully exploiting vectorization and parallelization
  techniques in a realistic experimental environment.
\end{abstract}

\section{Introduction}
\IEEEPARstart{T}{he} Large Hadron Collider (LHC) at CERN is the
highest energy collider ever constructed. It consists of two
counter-circulating proton beams made to interact in four locations
around a 17 mile ring straddling the border between Switzerland and
France. It is by some measures the largest man-made scientific device
on the planet. The goal of the LHC is to probe the basic building
blocks of matter and their interactions. In 2012, the Higgs boson was
discovered by the CMS and ATLAS collaborations.  Experimentally, we
collide proton beams at the center of our detectors and, by measuring
the energy and momentum of the escaping particles, infer the existence
of massive particles that were created and decayed in the pp collision
and measure those massive particles' properties.  In all cases, track
reconstruction, i.e., the determination of the trajectories of charged
particles, plays a key role in identifying particles and measuring
their charge and momentum. The track reconstruction as a whole is the
most computationally complex and time consuming, sensitive to
increased activity in the detector, and traditionally, least amenable
to parallelization.  The speed of reconstruction has a direct impact
on how much data can be stored from the 40 MHz collisions rate. The
most CPU-intensive part of the event selection in the online trigger
process is track reconstruction, to the point that it can only be
applied to a few \% of the input event rate. At the same time, the
tracker is the most precise instrument in CMS. Tracking is thus the
essential tool in making surgical decisions in the online selection to
optimally use the limited output bandwidth for interesting physics
events. The speed of track reconstruction software thus limits both
the total output rate of the experiments via the first cap, and the
surgical precision with which interesting events can be selected via
the second cap. Our research aims to lift those caps by vastly
speeding up the online tracking reconstruction.

\begin{figure}[!t]
  \centering
  \includegraphics[width=0.5\textwidth]{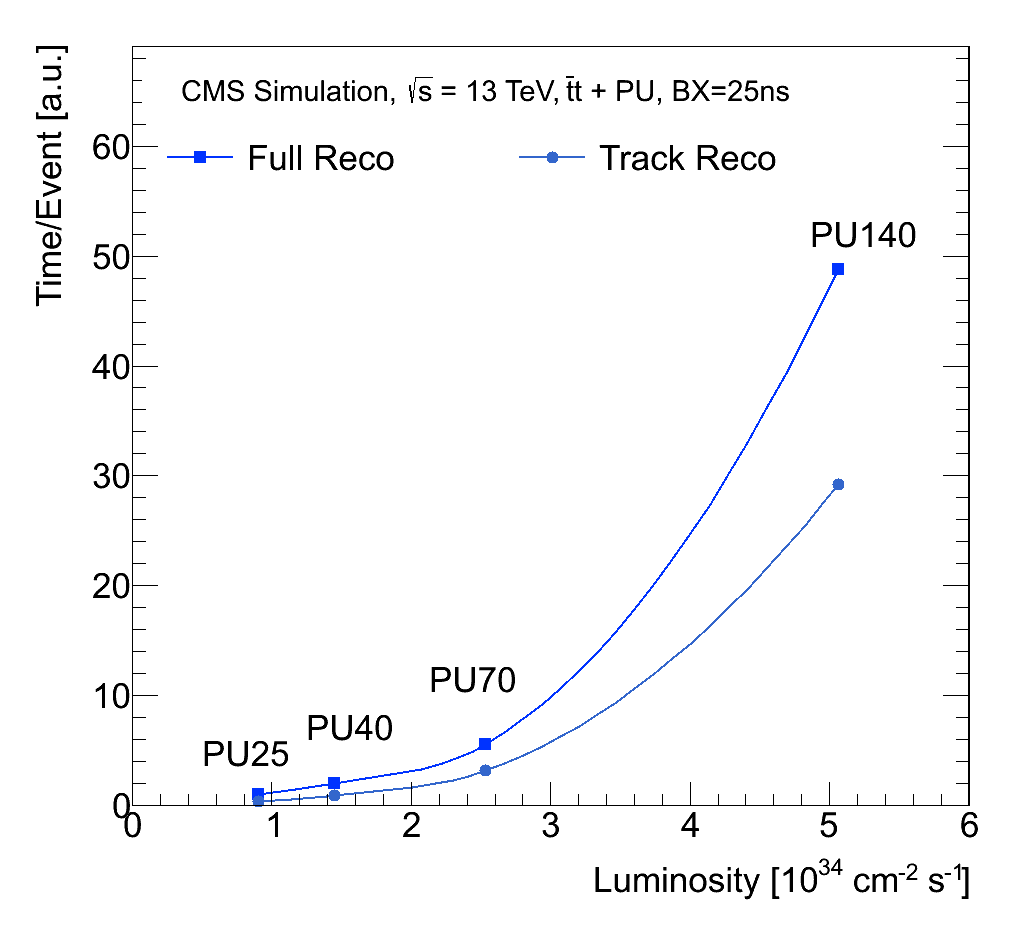}
  \caption{CPU time per event versus instantaneous luminosity, for
    both full reconstruction and the dominant tracking portion. PU25
    corresponds to the data taken during 2012, and PU140 corresponds to
    the HL-LHC era. The time of the reconstruction is dominated by
    track reconstruction. }
  \label{fig:pileup}
\end{figure}

The large time spent in track reconstruction will become even more
important in the HL-LHC era of the Large Hadron Collider, where the
increase in event rate will lead to an increase in detector occupancy
(``pile-up'', PU), leading to an exponential gain in time taken to do
track reconstruction, as can be seen in Fig.~\ref{fig:pileup}. In the
Figure, PU25 corresponds to the data taken during 2012, and PU140
corresponds to the low end of estimates for the HL-LHC era.
Clearly this research will become increasingly important during this era.

A correlated issue is the change in computing architectures in the
last decade. Around 2005, the computing processor market reached a
turning point: power density limitations in chips ended the long-time
trend of ever-increasing clock speeds, and our applications no longer
immediately run exponentially faster on subsequent generations of
processors. This is true even though the underlying transistor count
continues to increase per Moore's law. Increases in processing speed
such as those required by the time increases in Fig.~\ref{fig:pileup}
will no longer come `for free' from faster processors. New processors
instead are aggregates of `small cores' that in toto still show large
performance gains from generation to generation, but their usage
requires a re-work of our software to exploit. The processors in question
include ARM, GPGPU and the Intel Xeon Phi; in this work we target the
Xeon Phi architecture.

\section{Kalman Filter Tracking}
To realize the new performance gains, a change is required to move
from the sequential applications of today to vectorized, parallelized
applications of tomorrow.  The algorithm we are targeting for
parallelization is a Kalman Filter (KF) based
algorithm~\cite{Fruhwirth:1987fm}. KF-based tracking algorithms are
widely used since they naturally include estimates of scattering along
the trajectory of the particle due to multiple scattering off massive
detectors. Other algorithms, more naturally suited to parallelization
and coming from the image processing community, have been explored by
others. These include Hough Transforms and Cellular Automata, among
others (see, for instance, ~\cite{HALYO1}.)
However, these are not the main algorithms in use at the LHC today,
whereas there is extensive understanding how KF algorithms perform. KF
algorithms have proven to be robust and perform well in the difficult
experimental environment of the LHC. By porting these algorithms to
parallel architectures, we aim to port this robust tool to new
architectures.  Past work by our group has shown progress in
sub-stages of the KF algorithm in simplified detectors (see, e.g. our
presentations at ACAT2014~\cite{acat2014} and
CHEP2015~\cite{chep2015}).
\begin{figure}[!t]
  \centering
  \includegraphics[width=0.5\textwidth]{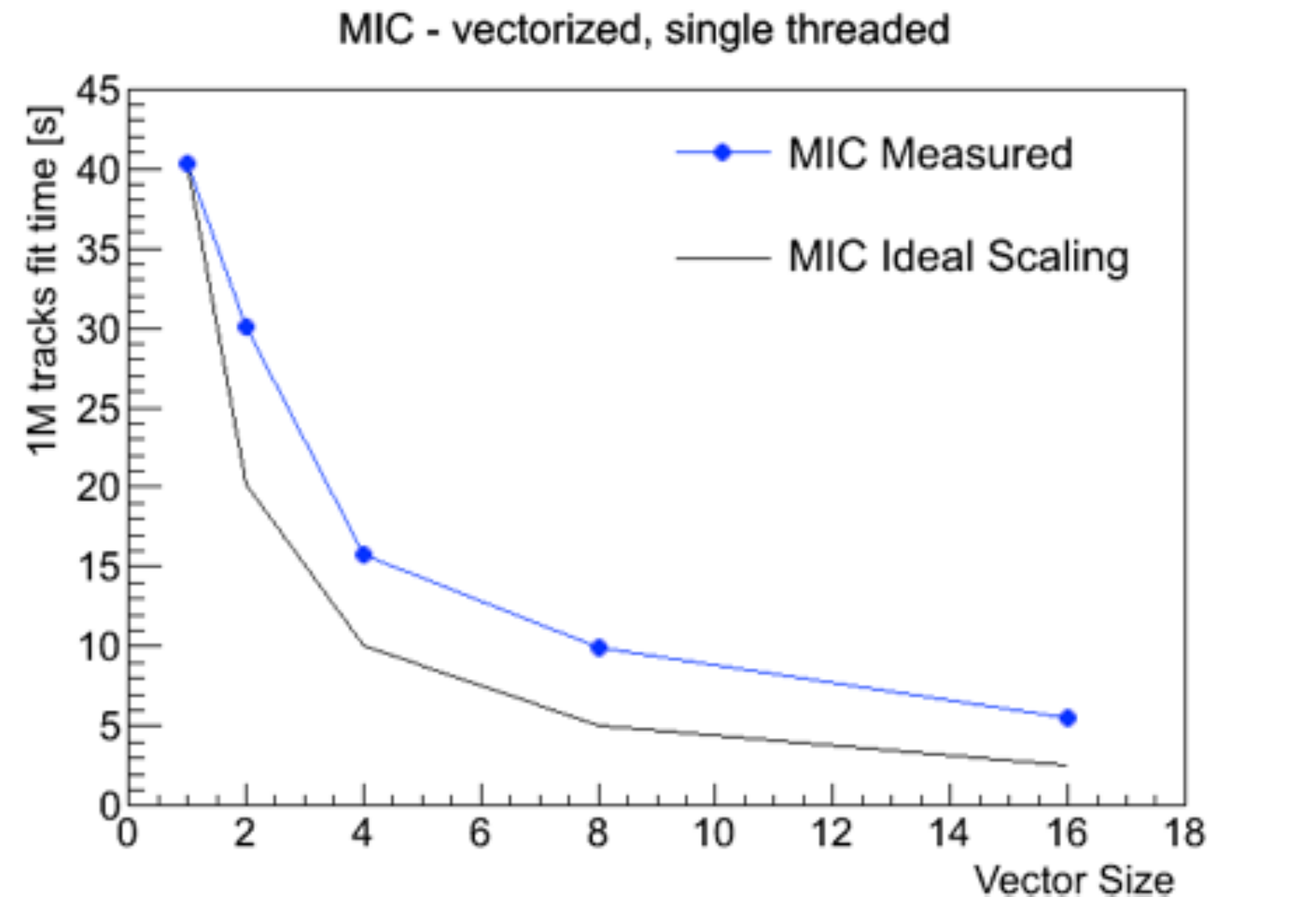}
  \caption{Track building part of the problem, where we decide which
    hits to group together as coming from the passage of a single
    charged particle, as a function of the increased usage of the
    processors' vector registers. }
  \label{fig:scaling}
\end{figure}
Fig.~\ref{fig:scaling} shows a result from
the track building part of the problem, where we decide which hits to
group together as coming from the passage of a single charged
particle, as a function of the increased usage of the processors'
vector registers. The black line shows ``ideal behavior'' - perfect
scaling, and the blue line shows our measured results. We observe a
significant speedup compared to the baseline, but still room for
improvement with respect to the ideal behavior. All results are for
the Xeon Phi. We have now implemented an end-to-end algorithm with a
semi-realistic detector model. This talk represents a status report.

% copied from ACAT14 paper
\subsection{Optimized Matrix Library \matriplex}
The computational problem of Kalman Filter-based tracking consists of
a sequence of matrix operations on matrices of sizes from
$N\times{}N = 3\times{}3$ up to $N\times{}N = 6\times{}6$. To allow
maximum flexibility for exploring SIMD operations on small-dimensional
matrices, and to decouple the specific computations from the high
level algorithm, we have developed a new matrix library,
\matriplex. The \matriplex memory layout is optimized for the loading
of vector registers for SIMD operations on a set of
matrices. \matriplex includes a code generator for generation of
optimized matrix operations supporting symmetric matrices and
on-the-fly matrix transposition. Patterns of elements which are known
by construction to be zero or one can be specified, and the resulting
generated code will be optimized accordingly to reduce unnecessary
register loads and arithmetic operations. The generated code can be
either standard \cplusplus or simple intrinsic macros that can be
easily mapped to architecture-specific intrinsic functions.
% \begin{itemize}
% \item New approach for a matrix library optimized for vectorization of
%   operations on small matrices 
% \item Matrix major representation: trivial loading of vector registers
%   for SIMD operations on a set of matrices. 
% \item Code generator for optimized matrix operations, supporting
%   symmetric matrices, transposes and known 0 and 1 elements. 
% \end{itemize}

\section{Current Status}
We have performed extensive studies of the performance of our
software. Both the physics performance and the code performance were
examined. For the latter, we used the \textsc{Intel
  VTune}~\cite{vtune} suite of tools to identify bottlenecks and
understand the impacts of our optimization attempts. In particular, as
can be expected, we determined that memory management is of critical
importance.  To this effect we describe below several studies to
optimize memory performance, and discuss the results of these studies.
\subsection{Cloning Engine Studies}
\begin{table*}[!t]
  \centering
  \begin{tabular}{|l|r|r|r|r|r|r|r|}
    \hline
&       VU01 Xeon &   VU08 Xeon &    speedup [\%]& VU01
                                                   Xeon Phi
    & VU16 Xeon Phi &      speedup [\%] \\
   & \multicolumn{2}{|c|}{s/10 events} & &\multicolumn{2}{|c|}{s/10
                                         events} & \\
    \hline
original &      17.4    &12.4&  29.0&   94.31&  70.76&  25 \\
cloning engine  & 17.87 &8.2   &  54  &    93   &   50 &     46 \\
threaded cloning engine&    11.6  &  6.13  &  47   &   64.5   & 38& 41 \\
\hline
speedup [\%]&   35    &  25 &     -      & 31 &     24 &      - \\
\hline
  \end{tabular}
  \caption{Performance Improvements from the cloning engine (CE) studies.  The
    first set of columns refers to traditional Xeon processors. The
    second set refers to Xeon Phi accelerators. VUXX refers to the
    vector unit size; maximum size is 8 in Xeon and 16 in Xeon
    Phi. The last row tabulates the speedup between the two CE models. }
  \label{tab:cloning}
\end{table*}
In Tab.~\ref{tab:cloning} we present results from our ``cloning
engine'' (CE) studies. In profiling our application it was observed
that a significant amount of time was spent in operations associated
with memory management. In our tracking algorithm, new data structures
have to be created when new track candidates are examined. Significant
time is also spent loading data into our local caches. In the CE
studies, we examine offloading the memory management work to a worker
process. The first strategy, labeled ``cloning engine'' in the Table,
gangs all memory operations together; speedup is observed by removing
redundant operations.  In the second strategy, labeled ``threaded
cloning engine'', the memory management task is delegated to a second
thread. We observe increased performance and vector unit efficiency
with CE studies compared to the original model and increased
performance for both Xeon and Xeon Phi using the threaded CE compared
to the non-threaded CE.

\subsection{Reduced Data Structures}
\begin{table}[!t]
  \centering
  \begin{tabular}{|l|r|r|}
    \hline
    &VU01 	&VU08  \\
    & [s/10 evt]	&  [s/10 evt] \\
    \hline
    original	& 17.6& 	12.3 \\
    original (r.d.f) &	13.9	 &8.7 \\
    cloning engine	& 19.3	& 9.7 \\
    cloning engine (r.d.f) &	18.3	&8.5 \\
    threaded cloning engine & 13.0	& 8.3 \\ 
    threaded cloning engine (r.d.f)	&13.8&	8.1 \\
    \hline
  \end{tabular}
  \caption{Performance scaling for various improvements. Measurements
    were performed on a Xeon CPU. VU01 and V08 refer to usage of
    vector unit of one and eight elements, respectively.  The acronym
    ``r.d.f'' refers to measurements with reduced data
    formats. Significant performance gains are seen by using reduced
    data formats. }
  \label{tab:scaling}
\end{table}
The size of the data structures used in our algorithm have a crucial
impact on the timing performance. In particular, the data structures
representing the energy deposits in the detector (the ``hits'') and
the reconstructed particle trajectories (the ``tracks'') must be
optimized in size to fit into the fastest cache memory.
Object-oriented data structures require more memory than arrays.
Therefore, we replaced the Kalman covariance matrices with basic
\textsc{C}-style arrays rather than \textsc{C++}-style classes. These
changes allowed us to reduce the size in memory of the track objects
by 20\% and the hit object by 40\%.  Tab.~\ref{tab:scaling} shows the
result of these studies.  The table only shows Xeon numbers; Xeon Phi
performance is similar.

In addition to moving to \textsc{C}-style data structures, we
optimized the contents of the data structures. Our original
implementation of these objects were designed for algorithmic ease of
use and contained data members not strictly needed for the track
reconstruction. For instance, Monte Carlo truth information,
\emph{i.e.}, information about how the trajectories were generated,
were stored in the hit objects. The track objects contained
(duplicate) copies of the hits themselves, rather than smaller
references to the hits in their original locations.  We rewrote the
data structures to optimize performance and maintain ease of use by
carefully considering what was needed for physics output only and
keeping auxiliary information in associated data structures.

We find significant speed-ups from the reduced data formated (labeled
``r.d.f'' in the table). We also find that with these changes, the
improvement of the CE studies is reduced; presumably since the amount
of memory churn overall, which the CE improves, has now been globally
reduced.

\subsection{Performance Scaling - parallelization}
\begin{figure*}[!t]
  \centering
  \includegraphics[width=1.0\textwidth]{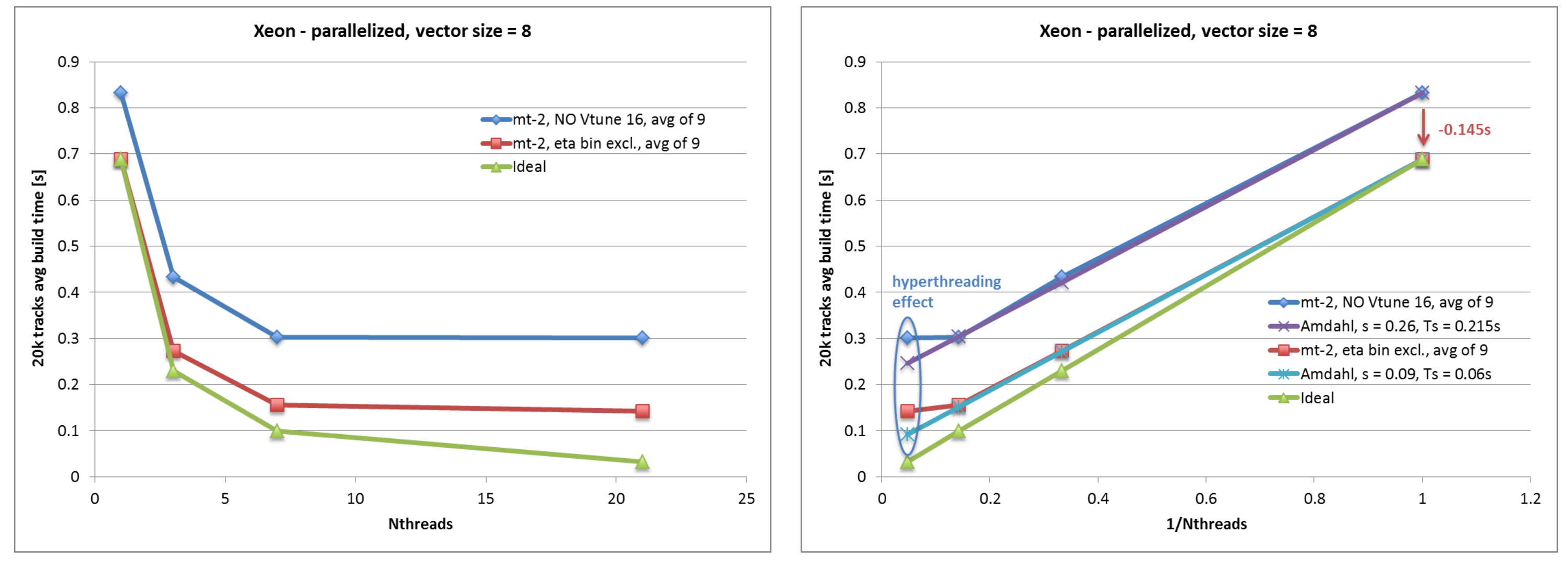}
  \caption{Performance scaling of the tracking code as a function of
    the number of threads, on the Xeon architecture. The plot on the
    right show the time to process 20,000 tracks as a function of the
    number of threads. The curve in green represent ideal scaling from
  the first data point. The blue and red curves show two different
  ways of distributing the data across threads. On the right the same
  data is plotted but now as a function of $1/n_\text{threads}$; this
  figure can be used to extract the serial fraction of the code.  }
  \label{fig:amdahl}
\end{figure*}
Figure~\ref{fig:amdahl} shows the parallelization performance scaling of
our code on the Xeon architecture. The plot on the right show the time
to process 20,000 tracks as a function of the number of threads. The
curve in green represent ideal scaling from the first data point. The
blue and red curves show two different ways of distributing the data
across threads. On the right the same data is plotted but now as a
function of $1/n_\text{threads}$; this figure can be used to extract
the serial fraction of the code. According to Amdahl's law, the time
spent on $n$ threads consists of a serial fraction $B$ and a parallel
fraction $(1-B)/n_\text{threads}$ multiplied by the time per thread, $T(1)$.
$$
T(n) = T(1)\left[ B + \dfrac{1}{n_\text{threads}} (1-B)\right]
$$
The $B$ parameter can be extracted from the data by fitting to this
function, and results are shown for different versions of our code.
In the Figure, a big improvement was found by optimizing how data
structures are re-initialized on each event, leading a reduction of
the serial fraction from 26\% to 9\%.  There is a significant residual
contribution to non-ideal scaling due to variation of occupancy within
threads: some threads simply take longer than others. In our group
there is work ongoing to define strategies for an efficient `next in
line' approach or a dynamic reallocation of thread resources to even
out timing across threads.
%%%% IEEE wants placement as !t
%\begin{figure}[!t]
%\centering
%%\includegraphics[width=3.5in]{myFigure}
%% where an .eps filename suffix will be assumed under latex, 
%% and a .pdf suffix will be assumed for pdflatex; or what has been declared
%% via \DeclareGraphicsExtensions.
%\caption{Daily abstract submission rate of the 2012 NSS-MIC. }
%\label{fig_sim}
%\end{figure}
%

\section{Conclusion}
We have made significant progress in parallelized and vectorized
Kalman-filter-based tracking R\&D on Xeon and Xeon Phi architectures.
We have developed a good understanding of bottlenecks and limitations
of our implementation.  New versions of the code are faster and
exhibit scaling closer to ideal performance. We are continuing to
pursue new ideas to further improve performance

Though it was not discussed in the talk, we have also developed tools
to process fully realistic data, with encouraging preliminary results.

The project is solid and promising; however, much work remains.

\section*{Acknowledgment}
This work was supported by the
U.S.  National Science Foundation.


\begin{thebibliography}{99}
\bibitem{Fruhwirth:1987fm}
R.~Fruehwirth, ``{Application of Kalman filtering to track and vertex
  fitting},''
\href{http://dx.doi.org/10.1016/0168-9002(87)90887-4}{{\em Nucl.~Instrum.~Meth.
  A} {\bfseries 262} (1987) 444--450}.
% \bibitem{GPU1}
% V.~Halyo, P.~LeGresley, and P.~Lujan, ``{Massively Parallel Computing and the
%   Search for Jets and Black Holes at the LHC},''
%   \href{http://dx.doi.org/10.1016/j.nima.2014.01.038}{{\em Nucl.~Instrum.~Meth.
%   A} {\bfseries 744} (2014) 54--60},
% \href{http://arxiv.org/abs/1309.6275}{{\ttfamily arXiv:1309.6275
%   [physics.comp-ph]}}.
% %%CITATION = ARXIV:1309.6275;%%.

% \bibitem{GPU2}
% V.~Halyo, A.~Hunt, P.~Jindal, P.~LeGresley, and P.~Lujan, ``{GPU Enhancement of
%   the Trigger to Extend Physics Reach at the LHC},''
%   \href{http://dx.doi.org/10.1088/1748-0221/8/10/P10005}{{\em JINST} {\bfseries
%   8} (2013) P10005},
% \href{http://arxiv.org/abs/1305.4855}{{\ttfamily arXiv:1305.4855
%   [physics.ins-det]}}.
% %%CITATION = ARXIV:1305.4855;%%.

\bibitem{HALYO1}
V.~Halyo, P.~LeGresley, P.~Lujan, V.~Karpusenko, and A.~Vladimirov, ``{First
  evaluation of the CPU, GPGPU and MIC architectures for real time particle
  tracking based on Hough transform at the LHC},'' {\em Journal of
  Instrumentation} {\bfseries 9} no.~04, (2014) P04005.
  \url{http://stacks.iop.org/1748-0221/9/i=04/a=P04005}.


\bibitem{acat2014} G.~Cerati \emph{et al.}, Traditional Track with Kalman Filter on Parallel Architectures, Proceedings of ACAT 2014, arXiv:1409.8213 (2014)
\bibitem{chep2015} G.~Cerati \emph{et al.}, Kalman Filter Tracking on
  Parallel Architectures, Proceedings of CHEP 2015, arXiv:1505.04540
  (2015).

\bibitem{vtune} See \url{https://software.intel.com/en-us/intel-vtune-amplifier-xe}.
\end{thebibliography}
\end{document}